\documentstyle[prb,multicol,aps,epsfig]{revtex}
\renewcommand{\narrowtext}{\begin{multicols}{2} \global\columnwidth20.5pc}
  \renewcommand{\widetext}{\end{multicols} \global\columnwidth42.5pc}

\begin{document}
\draft
\title{
The Signs of Quantum Dot-Lead Matrix Elements:\\ The Effect on Transport vs. Spectral Properties}

\author{Alessandro Silva, Yuval Oreg\cite{Incumbent} and Yuval Gefen.}
\address{Dept. of Condensed Matter Physics, The Weizmann Institute of Science, 76100
Rehovot, Israel.}

\maketitle
\begin{abstract}
{ A small quantum dot coupled to two external leads is considered.
Different signs of the dot-leads coupling matrix elements give rise to
qualitatively different behavior of  physical observables such as the conductance,
the phase of the transmission amplitude and the differential capacitance
of the dot. 
For certain relative signs the conductance may vanish
at values of the gate
potential, where the spectral density is maximal. Zeroes of the
conductance are robust against increasing the dot-lead
coupling. They are associated with abrupt phase lapses 
in the transmission phase whose width vanishes as the square of the
temperature. We carefully distinguish between phase lapses of
$-\pi$ and phase anti-lapses of $\pi$.}
\end{abstract}


\vspace{0.5cm}

\narrowtext
\section{Introduction}
Two interesting directions in the study of quantum dots (QDs) have
emerged in recent years. First, it has become clear that as the 
dot-lead coupling is increased, the effect of the Coulomb Blockade is
greatly suppressed, yet does not altogether disappear. Works
discussing the physics of strongly coupled QDs include
Refs.~\onlinecite{Matveev,Nazarov,Kamenev} (see also
Ref.~\onlinecite{Aleiner}). Second, experiments addressing the \it
phase \rm of the electron transmitted through a QD [employing an
Aharonov-Bohm (AB) set-up] revealed interesting and often
intriguing physics~\cite{Yacobi}.

Of the large number of theoretical works that followed, a few have addressed
(either explicitly or implicitly) the role of the signs of the dot-lead coupling
matrix elements~\cite{Baltin1,Oreg1}. No systematic study of the effect of the magnitude
and the relative sign of these coupling matrix elements on a number of physical
quantities, such as linear conductance, the spectral density and the transmission
phase has been carried out to date.

A small QD (or a small electron droplet), where the electron
spectrum is discrete, is a possible nanoscale laboratory for the
study of interference. [The discreteness of the spectrum of a
small dot is relevant when the mean level spacing $\bar{\Delta}
\gg \Gamma, k_B T$, where $\Gamma$ is the characteristic strength of the
coupling to the leads (cf. Eq.(\ref{Gamma}) below) and $T$ is the
temperature.] Indeed, in such systems there are various paths which
interfere and contribute to the conductance. These correspond to
different ways to traverse the QD, taking advantage of the various 
single particle levels. An efficient way to probe the effect of such
interference on the transmission amplitude 
through the QD (magnitude and phase)
is to embed the QD in one arm of an AB interferometer.
 
In a typical experimental set up the QD is connected to  source
and drain leads via tunneling barriers (or diffusive contacts)
with a typical coupling strength $\Gamma$, controlled through
additional gates. A ``plunger gate''  coupled
electrostatically to the QD controls the number of electrons in
the latter~\cite{Kouwenhoven}. As the potential $V_{ g}$ of the
plunger gate is increased, electrons are pulled into the dot. When
the coupling to the leads $\Gamma$ is small ($\Gamma \ll \bar
\Delta$) distinct peaks in the source-drain linear conductance
occur at near-degeneracy points~\cite{Kouwenhoven}, {\it i.e.},
when the energy of the dot with $N$ electrons is equal to the
energy of the dot with $N+1$ electrons (the Coulomb
peaks). The peaks in conductance  nearly coincide with the maxima
of the derivative of the mean number of electrons in the QD with 
respect to $V_g$. The
separation between the Coulomb peaks is mainly dictated by the
average charging energy of the dot, $U$.

The main goal of the present study is
to investigate the effect of the signs of the dot-lead coupling
matrix elements. The underlying physics is related to the interference among
different transmission amplitudes (say, from the left lead to the right lead),
describing different traversal paths through different single particle levels.
We analyze how this affects various physical quantities, such as the linear
conductance through the QD, the phase of the transmission amplitude as
measured by an AB interference experiment,
the spectral density of the QD, and the differential capacitance.

Reporting here the first part of our project, and attempting to
simplify the problem (ignoring some of the complex but interesting
ingredients), we consider  spin-less electrons and ignore at this
point electron-electron interactions. We also model the spectrum
of the small dot by two single particle levels. In a subsequent
work we will address the issue of an interacting QD. Below (see
Sect.~\ref{final}) we briefly comment on the relevance of studying
such a toy model for the sake of gaining insight into the physics
of ``real life'' QD.

We find that in addition to the strength of the coupling,
$\Gamma$, the key to understand the physics of the relevant physical
observables is the \it relative phase \rm of the coupling matrix
elements of consecutive orbital levels in the QD to the leads. 
This will be defined more accurately
below. Also, we find that the dependence of the conductance and
the differential capacitance (or the spectral density) as a
function of $V_g$ are complementary. For example, when the two consecutive
levels are \it in phase \rm ({\it i.e.}, the signs of the
respective couplings of level $1$ and level $2$ are identical),
the conductance exhibits two peaks whose positions are unaffected
by ~$\Gamma$. By contrast, the two peaks of
the spectral density (hence the differential capacitance) approach
each other as $\Gamma$ is increased, eventually merging into a single
peak. The complementarity is summarized in Table.~\ref{Table} below.

It should be noted that the mean number of electrons on the dot depends only
on the energy spectrum of the dot and the level width (given by
diagonal matrix elements of the single electron propagator). By
contrast, the conductance depends also on the actual value of the
eigenfunction of the system (including off-diagonal matrix
elements of the single electron propagator~\cite{Meir}). Hence it
is not surprising that over a certain range of values of $\Gamma$
the conductance and the differential capacitance exhibit
qualitatively different behavior~\cite{Kaminski} as a function of
$V_g$.

Another quantity which receives much attention here is the
phase of the total transmission amplitude. When the signs
of the respective couplings of the two dot's levels are identical
(the \it in-phase \rm scenario), it turns out that there is a value of
$V_g$ (corresponding to the ``conductance valley'' between the
two conductance peaks) for which the zero temperature conductance 
vanishes. This vanishing of the conductance is associated with a lapse 
of the transmission phase. It turns out (by carefully including 
the off-diagonal Green's functions
into the calculation of the transmission amplitude, see Appendix.~\ref{offdiagonal}) 
that the width of the phase lapse vanishes (like the square of the temperature). 
We point out in our analysis how it is possible to distinguish a $-\pi$ 
phase lapse from a $+\pi$ one, an observation which, we believe, has a 
greater range of validity than the specific problem considered here.
Surprisingly, the width of the phase lapse is determined by the 
temperature, to be contrasted with the width of the conductance
peak (the latter is determined by ${\rm Max}[\Gamma,T]$).

The remainder of the paper is organized as follows: in
Sect.~\ref{Sect1} we discuss the relation between measurable
quantities characterizing a QD, such as the conductance, the
AB oscillatory part of the conductance, the mean number
of electrons, the transmission amplitude, its phase and the
spectral density of electrons. In Sect.~\ref{Sect2} we introduce
a toy model consisting of a two level QD attached to two leads. We
recall an expression for the transmission amplitude through the QD
as well as for the spectral density (derivation of the latter can
be found in Appendix.~\ref{AppA}). In Sect.~\ref{Sect3} we study
the \it in-phase \rm case (all four coupling matrix elements having
the same sign). In Sect.~\ref{Sect4} we consider the 
\it out-of-phase \rm case, where one of
the matrix elements has a sign opposite to the others.

Section \ref{final} includes some remarks concerning
the relevance of the present analysis to the interacting case
and the possible extension of our model to more than two levels.

\section{Measurable quantities}\label{Sect1}

Experimentally, the basic quantities characterizing transport
through a QD are the conductance, the AB oscillations
pattern when the QD is inserted in an AB interferometer
and the differential capacitance, describing how the number of electrons
occupying the QD varies as  function of $V_g$.

Since we consider here a model of independent electrons, it is
possible to relate all these quantities\cite{Meir,Gefen} to the
transmission amplitude $t(\omega)$ through the QD and the spectral
density $A(\omega)$. We assume throughout our analysis single-channel
transport through each participating lead.


\subsection{The Conductance $G$}
Information about the absolute value  of $t(\omega)$ close to the Fermi
level can be obtained
through the linear response conductance, given by 
\begin{equation}\label{conductance}
G=-\frac{e^2}{h}\int d\omega f^{\prime}(\omega)\left|t(\omega)
\right|^2,
\end{equation}
where $f(\omega)$ is the Fermi function.

\subsection{Phase of the transmission amplitude $\theta$}

It is also possible to measure the phase of the
transmission amplitude through the QD by inserting it in one arm
of an open geometry AB interferometer, shown schematically in
Fig.~\ref{Fig7}. 

{\begin{figure}[ht] \vspace{0.1cm} \epsfxsize=6cm
\centerline{\epsfbox{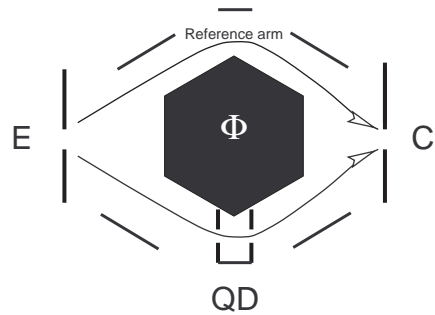}} \vspace*{0.5cm}
~\caption{ Schematic picture of an open geometry AB
interferometer. Electrons traverse the interferometer from the
emitter (E) to the collector (C). In one of the two arms a QD is
inserted; a magnetic flux $\Phi$ is enclosed in the area of
the interferometer. The two arrowed paths are
those that contribute the most to the conductance in the open geometry
configuration.} \label{Fig7}
\end{figure}}

One measures the dependence of the linear conductance
through the interferometer upon the enclosed flux, $\Phi$, 
and on the plunger gate voltage~$V_g$. It is possible to
show\cite{Oreg1,Oreg2,Gefen} that in this double-slit geometry
the flux-dependent oscillatory component of the conductance is
given by
\begin{equation}\label{ABosc}
G_{AB} \propto 2\; {\rm Re}\left[t^{*}_{\rm ref} \left(\int d\omega \;
\left(-f^{\prime}(\omega)\right) t(\omega)\right) e^{2\pi i
{\Phi/\Phi_0}}\right],
\end{equation}
where $t_{\rm ref}$ is the transmission amplitude through the
reference arm, assumed to be $V_g$ independent 
($t_{\rm ref}$ is taken energy independent as well);
${\Phi_0}=hc/e$ is the flux quantum. It is thus possible to
extract information about the temperature weighted phase of the
transmission amplitude through the QD
\begin{equation}\label{phase}
\theta(T)=\arg\left[ -\int d\omega \; f^{\prime}(\omega)
t(\omega)\right].
\end{equation}
In practice this is done by recording the AB oscillations as a
function of ${{\Phi}}$ for several values of the parameter~$V_g$.
The phase evolution (as a function of $V_g$) can be
extracted\cite{Yacobi} from the relative phase shift  of the
various curves.

We note that the ``transmission phase'' so measured may not
reflect the actual transmission phase through the QD, but might be
affected by multiple reflection paths, reflection from any of the
terminals of the interferometer and deviations from
unitarity~\cite{Ora,Feldman}.
\subsection{Differential Capacitance Measurements.}
The spectral density $A(\omega)$ represents the local
density of states in the QD at energy $\omega$ and is formally 
proportional to the imaginary part of the trace of the dot retarded
Green's function matrix (see Eq.~\ref{DOS2}). One can relate
the spectral density to the average number of electrons occupying
the QD through the expression
\begin{equation}
N=\int_{-\infty}^{\infty}\frac{d\omega}{2\pi} f(\omega)\;A(\omega),
\end{equation}
and study how $N$ varies as a function of $V_g$. In this case
the measured quantity is the differential capacitance
\begin{equation}\label{cap}
C(V_g)=\frac{d\;eN}{d\;V_g}=
e\;\int_{-\infty}^{\infty}\frac{d\omega}{2\pi}
f(\omega)\;\frac{d\;A(\omega)}{d\;V_g},
\end{equation}
where $e$ is the electronic charge.
The differential capacitance 
can be measured by means of a detector sensitive to the QD charge
\cite{Capacitance,Sprinzak}.

\section{A Toy Model Hamiltonian}\label{Sect2}
We shall consider the Hamiltonian
\begin{eqnarray}\label{Ham}
H&=&\sum \epsilon_{k,\alpha} \; c^{\dagger}_{k,\alpha}
c_{k,\alpha}+ \sum_j \epsilon_j
d^{\dagger}_jd_j+ \nonumber \\
&&+\sum_{k,\alpha,j} \left[V_{\alpha,j}c^{\dagger}_{k,\alpha}d_{j}+h.c.\right],
\end{eqnarray}
where the operators $c_{k,\alpha}$ refer to electronic states in
the leads ($\alpha=L,R$) and the operators $d_j$ describe the
quantum dot levels ($j=1,2$). In order to simplify our discussion,
we will assume the dot-lead couplings $V_{\alpha,j}$ to all have
the same magnitude but possibly different phases. It is
straightforward to see that three of these phases can be gauged
out~\cite{notephase}, absorbing them in the definitions of the operators
$c_{k,\alpha}$ and $d_{j}$. Hence, we choose~\cite{noteV}
$V_{L,1}=V_{R,1}=V_{L,2}=V$ and $V_{R,2}=e^{i\varphi}V$. When time
reversal symmetry is present, the QD's wave functions can be
chosen real, whereby, upon an appropriate gauge of the lead wave
functions, the value of the relative phase $\varphi$ is either $0$ or
$\pi$, in other words $s\equiv e^{i \varphi}= \pm 1$.

We are interested in calculating the transport properties of this
model as a function of a plunger gate voltage. This can be done
setting $\epsilon_{1,2}=- {\epsilon}\pm\Delta/2$, where
$\Delta=\epsilon_1-\epsilon_2$, and studying
the behavior of the system as a function of
\begin{equation}
 {\epsilon}(V_g)= {\epsilon}(V_g=0)+V_g.
\end{equation}
Other than the level spacing, the  scale coming
into play is the strength of the coupling to the
leads
\begin{equation}\label{Gamma}
\Gamma=2\pi\rho V^2,
\end{equation}
where $\rho$ is the density of states (DOS) of the leads.

Since the Hamiltonian is free (quadratic), the calculation
of both $t(\omega)$ and $A(\omega)$
is straightforward (see Appendix \ref{AppA}). One readily obtains
\begin{eqnarray}\label{transm}
t_{\pm}(\omega)&=&\frac{\Gamma}{D_\pm(\omega)}
\left[(\omega-\epsilon_1)\pm(\omega-\epsilon_2)\right],\\
\label{DOS} A_\pm(\omega)&=&\frac{2\Gamma}{\left|
D_\pm(\omega)\right|^2}
[2\omega^2-2\omega(\epsilon_1+\epsilon_2)\nonumber \\
&&+\epsilon_1^2+\epsilon_2^2+\Gamma^2(1\mp 1)]
,
\end{eqnarray}
where the denominator
$D_\pm(\omega)\equiv (\omega-\epsilon_1)(\omega-\epsilon_2)+
i\Gamma(2\omega-\epsilon_1-\epsilon_2)-\Gamma^2 (1\mp1)/2$.

\section{The in-phase (\lowercase{$s=+1$}) case}\label{Sect3}

As first step in our analysis we consider the \it in-phase \rm 
case, $s=+1$ [upper sign in Eqs.~(\ref{transm}),~(\ref{DOS})] . 
Starting with the zero temperature limit, all transport 
properties are determined by the  value of $t(\omega)$ 
at the Fermi level, {\it i.e.}, at $\omega = \epsilon_F\equiv 0$. 
From Eq.~(\ref{transm}) the transmission amplitude at the Fermi level
can be written as
\begin{equation}\label{tresult}
t_{s=+}(0)\equiv t_+ =\frac{2\Gamma {\epsilon}}{
{\epsilon}^2-(\Delta/2)^2+2i\Gamma {\epsilon}}.
\end{equation}
We now discuss the implications to the conductance and its flux sensitive component.

\subsection{The conductance G and the transmission-phase $\theta$ for
$s=+1$}
 At zero temperature the Fermi function in Eq.~(\ref{conductance}) can be
substituted by a Dirac delta function and $G$ is given by
$G=(e^2/h)t_{+}$. The conductance $G$ is depicted in
Fig.~\ref{Fig1} for various values of the parameters.

The main feature to be noticed in Fig.~\ref{Fig1} is the
presence of an exact zero of the transmission probability between
the two peaks, resulting from the vanishing of the numerator in
Eq.~(\ref{tresult}) at $\epsilon=0$. Physically that zero can be
interpreted as the result of destructive interference between
paths (left to right) traversing levels 1 and 2 of 
the dot respectively (see Appendix.~\ref{offdiagonal}).
As we demonstrate below, the interference pattern is sensitive to the
relative sign $s$.

The existence of a zero in the transmission amplitude implies the
existence of an abrupt (without a scale) phase
lapse~\cite{Yacobi,Oreg1,Berkovits,Weidenmuller,Theory} of $-\pi$ in the
``conductance valley'' between the two conductance peaks (cf.
Fig.~\ref{Fig1c}).

In terms of the AB oscillations pattern ({\it e.g.}, the
conductance measured as a function of $\Phi$) this implies that as
$ {\epsilon}$ varies from $0^{-}$ to $0^{+}$ $G(\Phi)$ shifts
abruptly by half a period. Since at $T=0$ this shift is abrupt, it
is physically impossible to discuss its direction, {\it i.e.}, 
whether  the phase of the AB oscillations jumps by $-\pi$ (lapse)
or by $+\pi$ (anti-lapse). Interestingly, at finite temperatures
this ambiguity is resolved: the phase varies by 
$-\pi$ close to $\epsilon=0$ (a lapse).
{\setlength{\unitlength}{1cm}
\begin{figure}
\vglue 0.5cm \epsfxsize=0.98\hsize
\hspace{-0.5cm}
\begin{picture}(5,5)
   \epsffile{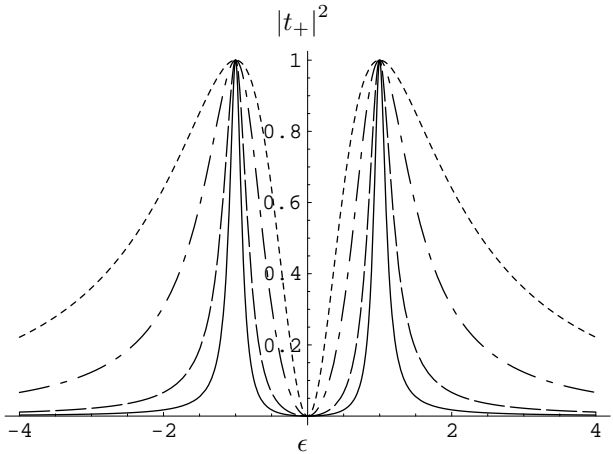}
   \put(-4.3,-0.1){\makebox(0,0)[b]{${ {\epsilon}}$}}
   \put(-4.3,5.5){\makebox(0,0)[b]{${\left| t_+ \right|^2}$}}
\end{picture}
\vspace{0.5cm} \caption{Transmission probability $\left| t_+
\right|^2$ at the Fermi energy of the leads vs. $ {\epsilon}$ for
$s=+$. Here $\epsilon_1-\epsilon_2=2$ and $\Gamma=0.1$ (full),
$0.2$ (dashed), $0.5$ (dash-dotted),$1$ (dotted). Notice that
(i)~at ${\epsilon}=0$ the contributions to the transmission 
through to the two levels (including off-diagonal elements of the
transmission matrix) add up
destructively leading to an exact zero of the transmission 
probability (see Appendix.~\ref{offdiagonal}); and
(ii)~the peak positions and maximal values (equal to 1) are insensitive
to $\Gamma$. The spectral density of the $s=-1$ case
($A_{-}$ in Fig.~\ref{Fig6}) exhibits a feature similar to the
latter for large values of $\Gamma$, 
while the spectral density $A_+$ (see Fig.~\ref{Fig1b})
has a maximum at $ \epsilon =0 $.}
\label{Fig1}
\end{figure}}
{\setlength{\unitlength}{1cm}
\begin{figure}
\vglue 0.5cm \epsfxsize=0.97\hsize
\begin{picture}(5,5)
\hspace{-0.5cm}
   \epsffile{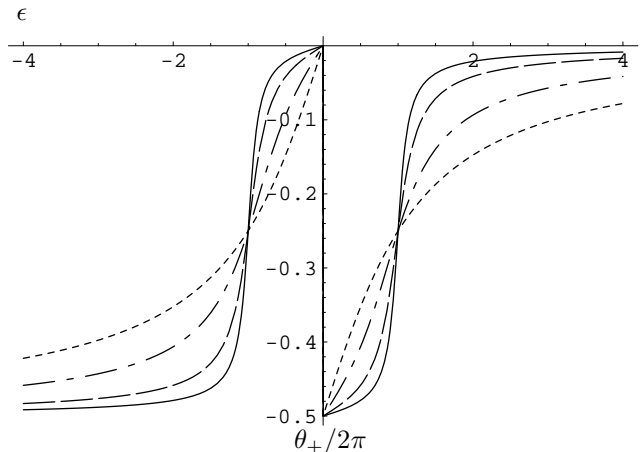}
   \put(-8.2,5.4){\makebox(0,0)[b]{${ {\epsilon}}$}}
   \put(-4.1,-0.3){\makebox(0,0)[b]{$\theta_+/2\pi$}}
\end{picture}
\vspace*{0.4cm} \caption{The transmission phase $\theta_+/2\pi$
at zero temperature vs.~${\epsilon}$ for $s=+$. Here,
$\epsilon_1-\epsilon_2=2$ and $\Gamma=0.1$~(full), $0.2$ (dashed),
$0.5$ (dash-dotted) and $1$ (dotted). Note that the destructive
interference of the transmission through the dot's levels leads to
the vanishing of the conductance at $ {\epsilon}=0$ which, in
turn, leads to an abrupt (without a scale) phase-lapse 
in the ``conductance valley''. 
As discussed below [cf. Eq.~(\ref{lapse1})], 
the width of the phase lapse at finite temperatures  is
$\propto \Gamma T^2/\Delta^2$, introducing a new nontrivial energy
scale. } \label{Fig1c}
\end{figure}}
This conclusion can obtained by noting that the trajectory in the
complex plane of $t_{+}$ [as a function of $ {\epsilon}(V_g)$] is
a closed curve tangential to the the abscissa at the origin 
(cf. the black line in Fig.~\ref{Complext}),
{\it i.e.}, $\rm{Im}[t(\omega)]\leq 0$. As $ {\epsilon}$  is
swept from $-\infty$ to $+\infty$, the transmission amplitude
performs  two full \em counterclockwise \rm revolutions, starting
from $t(-\infty)=[0^{-},0^{-}]$ and ending at
$t(\infty)=[0^{+},0^{-}]$. In the lower plot of
Fig.~\ref{Complext} we show how $t_+(  \epsilon)$ makes almost two
full revolutions around the circle when  $-4 <   \epsilon < 4$. It
starts (for $ \epsilon = -4$) at
$[{\rm Re}(t_+),{\rm Im}(t_+)] \approx [-0.3,-0.05]$, goes through
point~$(a)$ (cf. the upper plot of Fig.~\ref{Complext}) then
proceeds to point~$(b)$ at the origin and goes around
the circle through point~$(c)$.

For a fixed $  \epsilon $ at finite temperatures, one needs to
average over the sections of the curve close to 
$t_{+}({\epsilon})$, with the appropriate statistical weight
[cf. Eq.~(\ref{transmtemperature})]. The result
is a trajectory (shown schematically as a grey contour in
Fig.~\ref{Complext})  which does not include
the origin; the phase $\theta^{\vphantom 0}_+(T)$ evolves from
$-\pi$ to $0$ with a lapse of a finite width at $ {\epsilon}=0$.
We note that \it by considering a vanishingly small \rm (yet non-zero)
\it temperature, it is possible to determine \rm that the origin is not included
within the closed contour, hence \it the transmission phase lapses by \rm $-\pi$
(rather than by $+\pi$).

For $0<T\ll \Delta,\Gamma$, the evolution of the phase
$\theta^{\vphantom 0}_+$ for $ {\epsilon}\approx 0$ is well
approximated by (see Appendix \ref{AppB})
\begin{equation}\label{lapse}
\theta_+^{\vphantom 0}({\epsilon}) \simeq {\rm ArcTan}\left[
{\epsilon}/\lambda\right]-\frac{\pi}{2}.
\end{equation}
The width of the phase lapse is therefore given by a nontrivial
combination of $T$, $\Gamma$ and $\Delta$:
\begin{equation}\label{lapse1}
 \lambda \simeq (8 \pi^2/3)\;\Gamma\;T^2/\;\Delta^2.
\end{equation}
We note that this contrasts with the width of the conductance
peaks~\cite{Kouwenhoven}, $\lambda_{\rm{peak}} \sim \max[\Gamma,
T]$. Therefore, the quadratic dependence of $ \lambda$ on the $T$
leads at $T, \Gamma \ll \Delta$ to the inequality
$$
 \lambda \ll \lambda_{\rm{peak}} \mbox{ for } T, \Gamma \ll
\Delta.
$$
The smallness of ${\lambda}$ is in qualitative agreement with 
experimental observations~\cite{Yacobi}.
\begin{figure}[ht] \vglue 11cm \epsfxsize=0.93\hsize
\begin{picture}(5,5)
  \put(-14,0){\epsffile{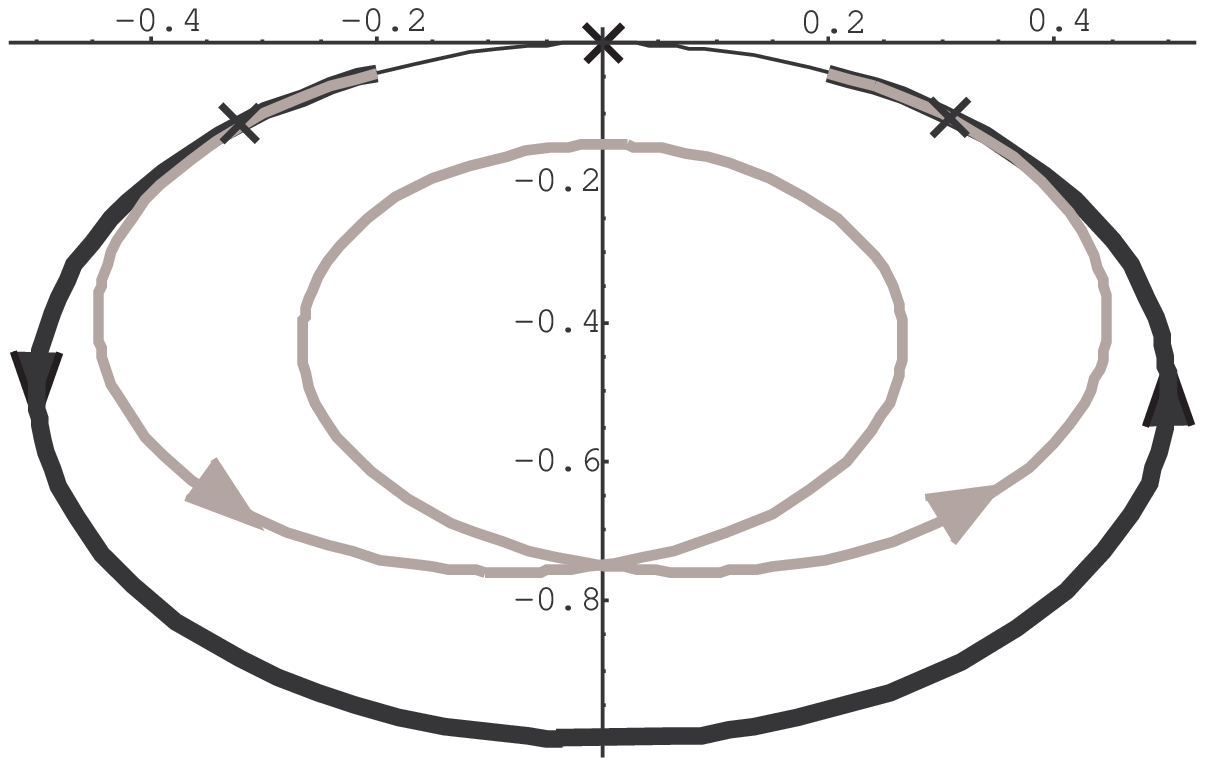}}
   \put(48,114){\makebox(0,0)[b]{ (a) }}
   \put(114,132){\makebox(0,0)[b]{(b)}}
   \put(181,114){\makebox(0,0)[b]{ (c) }}
  \put(114,-8){\makebox(0,0)[b]{${\rm Im}\; t_+$}}
   \put(221,114){\makebox(0,0)[b]{${{\rm Re}\; t_+}$}}
\end{picture}
\epsfxsize=0.925\hsize
\begin{picture}(5,5)
 \put(-10,174){\epsffile{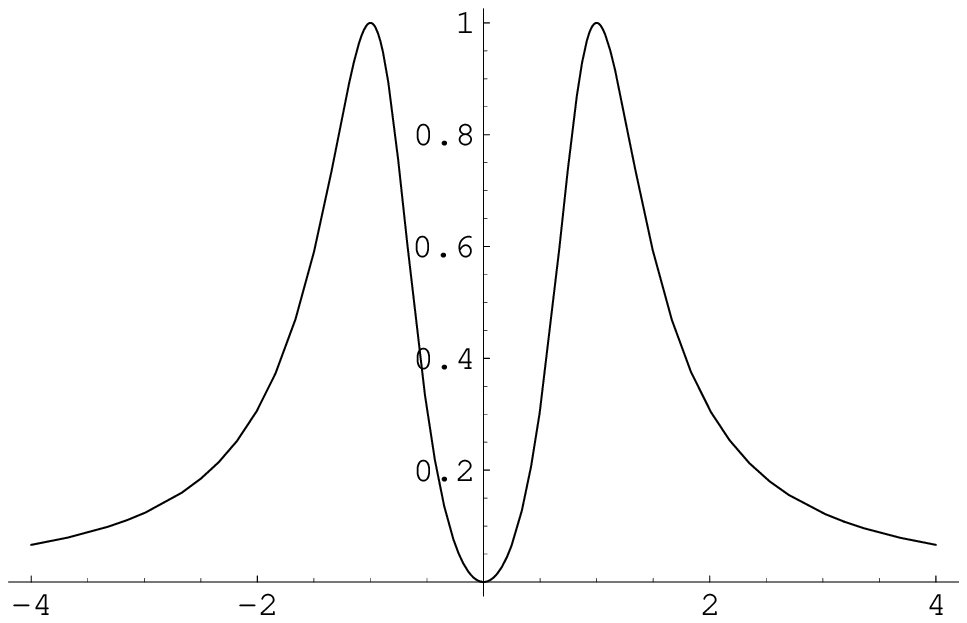}}
 \put(204,168){$ {\epsilon}$}
  \put(96,320){$ \small{\left| t_{+} \right|^2} $}
\put(30,172){\makebox(0,0)[b]{(a)}}
\put(30,181){\makebox(0,0)[b]{$\times$}}
 \put(75,172){\makebox(0,0)[b]{$$}}
 \put(133,172){\makebox(0,0)[b]{$$}}
 \put(104,172){\makebox(0,0)[b]{(b)}}
 \put(178,172){\makebox(0,0)[b]{(c)}}
  \put(104,181){\makebox(0,0)[b]{$\times$}}
 \put(178,181){\makebox(0,0)[b]{$\times$}}
\end{picture}
\vspace*{0.5cm} \caption{ Plot of $\left| t_+( {\epsilon})\right|^2$
vs. $ {\epsilon}$ for $\epsilon_1-\epsilon_2=2$ and
$\Gamma=0.5$ (upper plot) and of $t_+( {\epsilon})$ in the complex
plane (lower plot). At zero temperature the transmission
amplitude evolves along a circle in the lower part of the complex
plane ($\rm{Im}[t]\leq 0$). As $ {\epsilon}$ is increased, 
$t_+({\epsilon})$ evolves from point~(a) traversing the origin at 
${\epsilon}= 0$ [point~(b)] and winds around the circle again up to
point~(c). The black thick line in the lower plot denotes that portion of
the circle that is visited twice by $t_+(  \epsilon)$ as  $\epsilon$  
varies from $-4$ to $4$.
Upon temperature averaging $t_+( {\epsilon})$ evolves along
a contour enclosed in the zero temperature circle (grey line in the lower
plot at $T=0.2$). As a result, the finite temperature curve does not contain
the origin, and the phase lapse of Fig.~\ref{Fig1c} is smeared.
It is then easily concluded that as $T \rightarrow 0$ the transmission
phase lapses by $-\pi$.
\label{Complext}}
\end{figure}
As seen from  Fig.~\ref{Fig1}, the presence of a zero of the
transmission amplitude implies that $G( {\epsilon})$ has a
peak-valley-peak structure for \it all \rm values of $\Gamma$. The width
(but not the depth) of the conductance valley shrinks as $\Gamma$
is increased, and the value of the transmission probability at the
peaks is always equal to $1$.

\subsection{Differential Capacitance $C$ for $s=+1$}

The features seen in the conductance are now contrasted with the
behavior of the spectral density, $A_+$ (at the Fermi energy) as a
function of $ {\epsilon}$, depicted in Fig.~\ref{Fig1b}. Unlike
the destructive interference that leads to the vanishing of the
conductance at $  \epsilon =0$, the two peaks in the spectral
density tend to \em merge \rm into a single peak as $\Gamma$
increases.

Formally the difference between these two quantities is seen in
Eq.~(\ref{tgeneral}) for the transmission and Eq.~(\ref{DOS2}) for
the spectral density. While the former depends on all the elements
of the Green function matrix, including the off-diagonal ones, the
latter contains information on its trace only. Physically, the
conductance depends directly on the size and signs of the
couplings to the leads, as it describes the transfer of an
electron from the left lead to the right one through both levels of the QD.
The spectral density, on the other hand, depends on the
couplings to the leads only through the
modification of the position and width of the bare levels.

{\setlength{\unitlength}{1cm}
\begin{figure}
\vglue 0.5cm \epsfxsize=1\hsize
\hspace*{-0.5cm}
\begin{picture}(5,5)
\epsffile{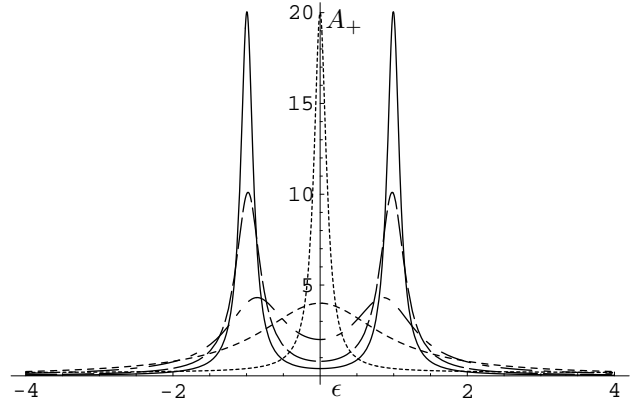}
   \put(-4.1,0.1){\makebox(0,0)[b]{${ {\epsilon}}$}}
   \put(-4,5){\makebox(0,0)[b]{${ A_+}$}}
\end{picture}
\vspace{0.3cm}\caption{The spectral density $A_+$ [given by
Eq.~(\ref{DOS}) and related to the differential capacitance via
Eq.~(\protect{\ref{eq:CA}})], vs. $ {\epsilon}$ for $s=+1$. Here
$\epsilon_1-\epsilon_2=2$ and $\Gamma=0.1$ (full), $0.2$ (dashed),
$0.5$ (dash-dotted),$1$ (dotted), and $5$ (small dots). Note that
unlike the conductance $|t_+|^2$ depicted in
Fig.~\protect{\ref{Fig1}}, here there is a peak emerging 
at $ \epsilon=0$ for large values of $\Gamma$.}
\label{Fig1b}
\end{figure}}

To gain some insight into the behavior of the spectral density, it is useful to
rewrite it as the sum of two Lorentzians
\begin{equation}\label{DOS1}
A_+(\omega)= -2 \;{\rm
Im}\left[\frac{1}{\omega-\omega_{e}}+\frac{1}{\omega-\omega_{o}}\right],
\end{equation}
where the poles are given by
\begin{equation}\label{poles}
\omega_{e(o)}=- {\epsilon}-i\Gamma \pm
\sqrt{(\Delta/2)^2-\Gamma^2}.
\end{equation}
The expression for $A_+(\omega)$ shows that for 
$\Gamma \ll \Delta/2$ the two Lorentzians are centered at the positions of the
original levels ($\epsilon_{1,2}$) and have each a  width
$\Gamma$. As $\Gamma$ exceeds $\Delta/2$ the picture is
drastically revised --- in this limit the two peaks of the
spectral density \it merge \rm to form two peaks  centered at 
$-{\epsilon}$ with different widths 
$\Gamma_{e(o)}=\Gamma \pm \sqrt{\Gamma^2-(\Delta/2)^2}$. 
Indeed, while one of these two
peaks of $A_+$ broadens as $\Gamma$ is increased, the second one
becomes increasingly sharper~\cite{Levit} (cf. Fig.~\ref{Fig1b}). This behavior
is directly reflected in the differential capacitance (as a
function of $V_g$). At zero temperature,  the
differential capacitance [Eq.(\ref{cap})] of our toy model 
is given by
\begin{equation}
\label{eq:CA} C(V_g)=e\;\frac{A(\omega=0)}{2\pi}.
\end{equation}

%
The fact that the transmission zero and the spectral \it merging
\rm occur concomitantly can be understood through a simple change
of variables. Let us perform a canonical transformation to even
and odd combinations of the dot's operators
\begin{eqnarray}\label{symm}
d_{e}=\frac{d_1+d_2}{\sqrt{2}},\;\;d_{o}=\frac{d_1-d_2}{\sqrt{2}}.
\end{eqnarray}
Substituting in the Hamiltonian, Eq.(\ref{Ham}), we obtain
\begin{eqnarray}\label{Ham1}
H&=&\sum \epsilon_{k,i} \; c^{\dagger}_{k,i} c_{k,i}-
 {\epsilon}(d^{\dagger}_{o}d_{o}+
d^{\dagger}_{e}d_{e})+\frac{\Delta}{2} (d^{\dagger}_{e}d_{o}+h.c.)\nonumber \\
&&+\sum_{k}
V\left[(c^{\dagger}_{k,L}+c^{\dagger}_{k,R})d_{e}+h.c.\right].
\end{eqnarray}
In this representation, only one combination (even) is directly
coupled to the leads. However, since both the even and odd modes
are not eigenstates of the dot's Hamiltonian they are coupled by a
tunneling term, whose strength is $\Delta/2$.

Employing this transformation, we readily understand the behavior
of the spectral density for $\Gamma \gtrsim \Delta/2$. Indeed, in
this case the two concentric peaks of the spectral density $A_+$
essentially correspond to the even (broad peak) and odd (narrow
peak) combinations respectively. The reason why in this limit it
is particularly useful to stick to the even-odd basis is that the
escape time $1/\Gamma$ from the even combination to the leads is
shorter than the typical time for tunneling between the two modes
$2/\Delta$. Since the even mode is directly coupled to the leads
its width is larger than that of the
odd mode.

\section{Analysis of the case \lowercase{$s=-1$}}\label{Sect4}
As anticipated in the introduction, the behavior described above
is not universal but depends crucially on the relative sign of the
coupling constants, $s$. Indeed,  the qualitative features for
$s=-1$ are different from the ones described above for $s=+1$. In
this case the spectral density assumes the form [cf.
Eq.(\ref{DOS})]
\begin{equation}\label{Api}
A_{-}(\omega)=\frac{2\Gamma}{(\omega+
 {\epsilon}-\Delta)^2+\Gamma^2}+
\frac{2\Gamma}{(\omega+ {\epsilon}+\Delta)^2+\Gamma^2},
\end{equation}
while the transmission amplitude is given by [cf.
Eq.(\ref{transm})]
\begin{equation}\label{tpi}
t_-(\omega)=\Gamma \left[\frac{1}{\omega+
 {\epsilon}-\Delta+i\Gamma}-
\frac{1}{\omega+ {\epsilon}+\Delta+i\Gamma}\right].
\end{equation}
\subsection{The conductance $G$ and transmission phase $\theta$ for $s=-1$}
We first consider the transmission probability depicted in
Fig.~\ref{Fig4}. The features of the case $s=-1$ are markedly different from the
previous $s=+1$ case. First, the transmission probability $\left|
t_- \right|^2$ is always finite over the entire energy ($
{\epsilon}$) range, implying  the absence of a phase lapse.
Indeed, the phase evolves continuously from zero to $2\pi$ as $
{\epsilon}$ is swept across the two resonances.
{\setlength{\unitlength}{1cm}
\begin{figure}
\vglue 0.6cm \epsfxsize=1\hsize
\begin{picture}(5,5)
\hspace{-0.5cm}
 \epsffile{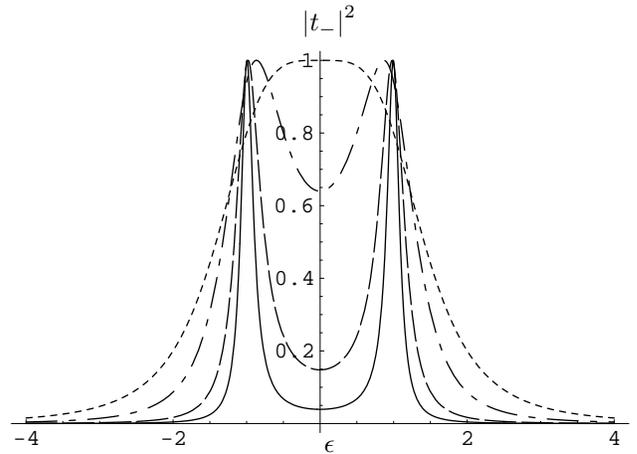}
   \put(-4.2,0.0){\makebox(0,0)[b]{${ {\epsilon}}$}}
   \put(-4.2,5.6){\makebox(0,0)[b]{${\left| t_- \right|^2}$}}
\end{picture}
\vspace{0.4cm}\caption{The transmission probability $\left| t_-
\right|^2$ at the Fermi energy vs. $ {\epsilon}$ for $s=-1$. Here,
$\Delta=\epsilon_1-\epsilon_2=2$ and $\Gamma=0.1$ (full), $0.2$
(dashed), $0.5$ (dash-dotted),$1$ (dotted). Notice that in
contrast to the zero at $  \epsilon =0$ for $s=+1$
(Fig.~\ref{Fig1}), here at large $\Gamma$ the conductance is
peaked at $  \epsilon =0$. In similarity to the
spectral density $A_+$ (Fig.~\ref{Fig1b}), as $\Gamma$
increases the conductance peaks approach each and subsequently
merge~\cite{noteYuval}.} \label{Fig4}
\end{figure}}
Secondly, the positions of the maxima of the transmission
probability $\left| t_- \right|^2$ shift as $\Gamma$ is increased
(contrary to the $s=+1$ scenario, Fig.~\ref{Fig1c}); they are
given by $\omega^{\pm}=- {\epsilon} \pm
\sqrt{(\Delta/2)^2-\Gamma^2}$ for $\Gamma \lesssim \Delta/2$. For
$\Gamma \gtrsim \Delta/2$ the two peaks merge~\cite{noteYuval} and are centered at
$- {\epsilon}$. This behavior is reminiscent of the one observed
in the spectral density in the $s=+1$ case, cf. Fig.~\ref{Fig1b}.
{\setlength{\unitlength}{1cm}
\begin{figure}
\vglue 0.5cm \epsfxsize=1\hsize
\hspace{-0.5cm}
\begin{picture}(5,5)
 \epsffile{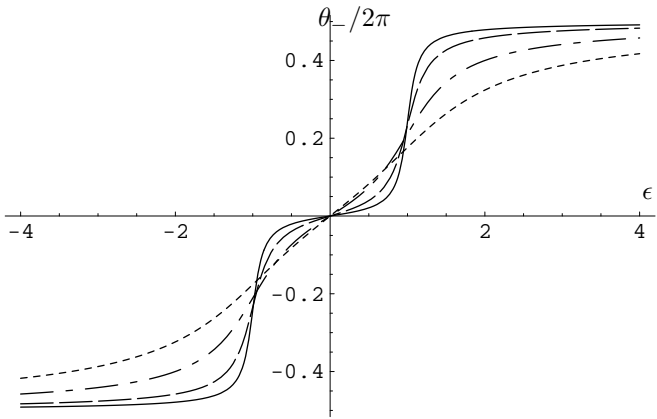}
   \put(-0.1,2.9){\makebox(0,0)[b]{${ {\epsilon}}$}}
   \put(-4,5.2){\makebox(0,0)[b]{${\theta_-/2\pi}$}}
\end{picture}
\vspace{0.5cm}\caption{The transmission phase $\theta_-/2\pi$ at
zero temperature vs. $ {\epsilon}$ for $s=+1$. Here,
$\Delta=\epsilon_1-\epsilon_2=2$ and $\Gamma=0.1$ (full), $0.2$
(dashed), $0.5$ (dash-dotted),$1$ (dotted). Notice that in
contrast to Fig.~\ref{Fig1} the phase evolves in the conductance
valley continuously from zero to $2 \pi$, hence no phase
lapse. } \label{Fig5}
\end{figure}}
It is quite interesting to note that for the present $s=-1$ scenario
the peaks in the spectral density do not shift (unlike the $s=+1$ case).
This is seen from Eq.(\ref{Api}) and Fig.~\ref{Fig6}.
{\setlength{\unitlength}{1cm}
\begin{figure}
\vglue 0.5cm \epsfxsize 1\hsize
\hspace{-0.5cm}
\begin{picture}(5,5)
\epsffile{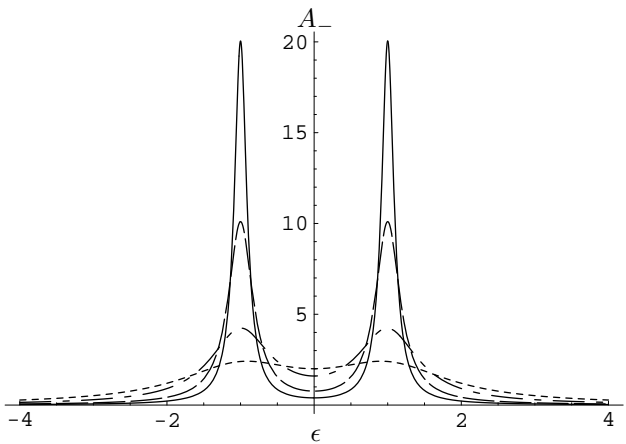}
   \put(-4.3,-0.1){\makebox(0,0)[b]{${ {\epsilon}}$}}
   \put(-4.3,5.4){\makebox(0,0)[b]{$A_-$}}
\end{picture}
\vspace{0.5cm} \caption{The spectral density $A_-$ at the Fermi
energy vs. $ {\epsilon}$ for $s=-1$. Here
$\Delta=\epsilon_1-\epsilon_2=2$ and $\Gamma=0.1$ (full),
$0.2$~(dashed), $0.5$ (dash-dotted), $1$ (dotted). Note that the
peak positions do not depend on the strength of the coupling to
the leads, $\Gamma$, similarly to the behavior of $\left| t_+
\right| ^2$ depicted in Fig.~\ref{Fig1}.} \label{Fig6}
\end{figure}}

A concise way to summarize the qualitative behavior of the cases
$s=\pm 1$ is by noting that $\left| t_{\pm}\right|^2$ and $A_{\pm}$
behave in a complementary manner as far as the peak position is 
concerned~\cite{noteYuval}. For both $\left| t_+ \right|^2$
and for $A_-$ the peak positions are insensitive to the magnitude
of $\Gamma$, while for both $\left| t_- \right|^2$ and for $A_+$
the peaks approach each other as $\Gamma$
is increased and finally they merge. The reader may consult
Table.~\ref{Table}.

\section{Brief summary and possible extensions.}\label{final}

The preceding analysis was restricted to a toy model of a QD with
two adjacent levels. The qualitative features described above
would survive the extension of our model to $N>2$ levels. As
$\Gamma$ is increased towards $\Delta$,
the crucial parameter determining the behavior of the conductance,
the transmission phase and the spectral density is the relative
sign of the coupling matrix elements associated with consecutive levels.
As an example
Fig.~\ref{Fig8} and Fig.~\ref{Fig9} depict the transmission
probability $\left| t \right|^2$ and the transmission phase
$\theta$ respectively for a set of 7 levels, the first four 
``in-phase'' ($s=+1$) while the next three levels being 
``out of-phase'' ($s=-1$). Similarly to the two-level model, we observe
here zeroes of the transmission amplitude as well as phase lapses
($s=+$). The peak structure for $s=-1$ is blurred, indicating an
incipient merger of the peaks (cf. Fig.~\ref{Fig4}). 
{\setlength{\unitlength}{1cm}
\begin{figure}
\vglue 0.5cm \epsfxsize 1\hsize
\hspace{-0.5cm}
\begin{picture}(5,5)
 \epsffile{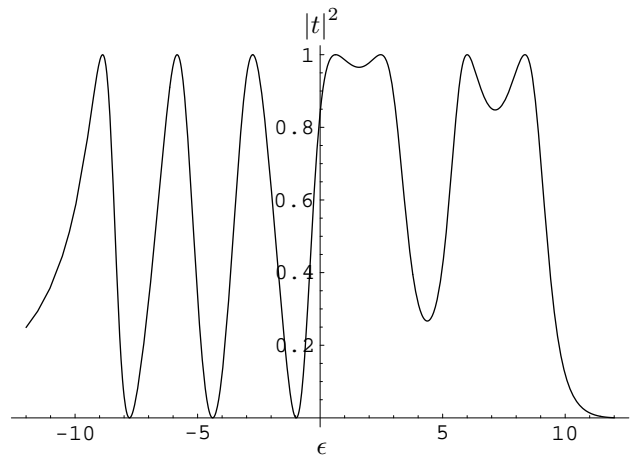}
   \put(-4.3,-0.1){\makebox(0,0)[b]{${ {\epsilon}}$}}
   \put(-4.3,5.5){\makebox(0,0)[b]{$\left| t \right|^2$}}
\end{picture}
\vspace{0.5cm} \caption{The transmission probability $\left| t
\right|^2$ at the Fermi level vs. $ {\epsilon}$ for 7 levels, with
level spacing $\Delta=3$. All couplings to the leads have the
same magnitude (for every level $\Gamma=\Gamma_L+\Gamma_R=1$).
However, the first four levels 
are in phase, while the last three (peaks on the right) have a
relative phase of $\pi$ between each consecutive pair. Note
that the ``conductance valley'' between, e.g., peaks 4 and 5 
almost disappears. All the
qualitative features described in the text for two levels are
observed also here.} \label{Fig8}
\end{figure}}
Once $\Gamma,T
\gg \Delta$ (for our noninteracting model) more than two
levels will contribute to the transmission amplitude  of electrons
at a certain energy.
In this case, for every value of the gate voltage, the details of
the couplings to the leads (magnitude and phase) of a set of $N
\simeq\Gamma/\Delta$ levels around the Fermi level 
determine the behavior of the system.
{\setlength{\unitlength}{1cm}
\begin{figure}
\vglue 0.2cm \epsfxsize 1\hsize
\hspace{-0.5cm}
\begin{picture}(5,5)
\epsffile{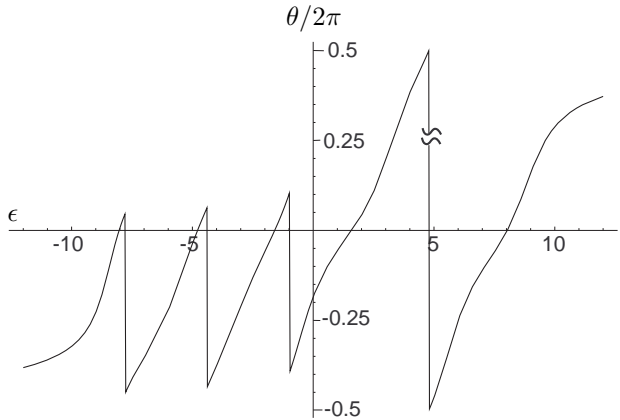}
   \put(-8.3,2.9){\makebox(0,0)[b]{${ {\epsilon}}$}}
   \put(-4.3,5.5){\makebox(0,0)[b]{$\theta/2\pi$}}
\end{picture}
\vspace{0.5cm} \caption{The transmission phase $\theta/2\pi$ vs.
$ {\epsilon}$ for 7 levels, with level spacing $\Delta=3$. All
couplings to the leads have the same absolute value (for every
level $\Gamma=1$) (cf. the caption of Fig.~\ref{Fig8}). Since the
first four levels are ``in phase'' ($s=+1$), a phase lapse of $-\pi$
is observed in the valleys between peaks (1,2), (2,3), and (3,4). 
For the remaining levels, being ``out of-phase'', the phase evolves continuously. The
discontinuous jump seen at $ {\epsilon}\simeq 5$ is due to the
fact that the phase is defined between $0$ and $2\pi$ and \it does
not \rm reflect any physical effect. } \label{Fig9}
\end{figure}}
The issue of electron-electron interaction deserves careful consideration,
beyond the scope of the present paper. A few remarks are nevertheless due.
One can account for the interaction on the level of a capacitive term, incorporating the
standard term
\begin{equation}
\hat{H}_{\rm int}=U n_1 n_2\nonumber
\end{equation}
($n_i=d^{\dagger}_id_i$) in the Hamiltonian [cf. Eq.(\ref{Ham})]. It
is possible to treat this interaction term within a
self-consistent Hartree scheme. This approximation is
justified\cite{noteint} when $\Gamma <\Delta$. In this case the
main relevant features of the model studied here, {\it i.e.}, the
qualitative differences between the two cases $s=+1$ and $s=-1$ as
well as the contrasting behavior of the conductance and the
spectral density, remain unchanged. In particular, for $s=+1$ the
conductance valley contains a zero and therefore the phase exhibits a
lapse in similitude to the noninteracting case. Within this scheme the main
effect of including the interactions is the replacement of the
bare levels $\epsilon_1,\epsilon_2$ by the self consistent Hartree
levels
\begin{equation}
\epsilon_{1}^{\prime}=\epsilon_{1}+U\langle n_2 \rangle\;\;\;;\;\;\;
\epsilon_{2}^{\prime}=\epsilon_{2}+U\langle n_1 \rangle.\nonumber
\end{equation}
It follows that the distance between consecutive conductance peaks
is then $\approx \Delta+U$.

In conclusion, we have analyzed a simple noninteracting toy model
describing a QD coupled to two leads and studied certain important
observables such as the transmission amplitude
through the QD (phase and magnitude), as well as the spectral
density. We have shown that the transmission probability and the
spectral density exhibit qualitatively different behavior as
function of the plunger voltage. These differences become
dramatically apparent when the coupling to the leads $\Gamma$
exceeds $\Delta$.

A crucially important parameter in our discussion is the relative phase
$\varphi$ between two adjacent levels in the QD.
The behavior of the various physical quantities depends strongly on
whether this phase is $\varphi=0$ ($s=+1$) or $\varphi=\pi$ ($s=-1$)(time reversal
symmetry is assumed). The salient features observed
in these two cases are summarized in Table I.
\widetext
{\begin{center}
\begin{table}
\begin{tabular}{|c|c|c|c|}
   &         & $s=+1$ & $s=-1$  \\ \hline
  $\;G$:& Conductance        & peaks do not shift & peaks merge at $\Gamma > \Delta/2$ \\ \hline
  $\;\theta$: & Transmission phase  & a sharp phase lapse of width $\propto \Gamma (T/\Delta)^2$ & no phase lapse\\ \hline
  $\;C$: & Differential capacitance      & peaks merge at $\Gamma > \Delta/2$ & peaks do not shift \\
\end{tabular}
\caption{\label{Table}Notice the complementarity  in the
qualitative features of the differential capacitance and the
conductance for $s=+1$ and $s=-1$.}
\end{table}
\end{center}}
\narrowtext

\section{Acknowledgment}

We would like to thank S.~Levit, Y.~Imry, M.~Heiblum, M.~Schechter,
J.~K\"{o}nig and B. Kubala for helpful discussions. Y.G. and A.S.
acknowledge the hospitality of ITP Santa Barbara where part of
this work was performed. This work has been supported by the Israel
Science Foundation of the Israel Academy grants, by the German-Israel
DIP grants, by the U.S.-Israel BSF, and by the German-Israel GIF.

\appendix
\section{ Derivation of the transmission amplitude and spectral density}\label{AppA}

In this appendix we sketch the derivation of the transmission
amplitude and the spectral density,
Eqs.(\ref{transm})~and~(\ref{DOS}). All physical quantities
discussed here can be expressed in terms of the QD Green's
functions ${\bf G}_{i,j}(t)\equiv -i \theta(t) \langle
{d_i(t),d^{\dagger}_j(0)}\rangle$. This Green's function matrix is
given by
\begin{equation}\label{green1}
{\bf G}(\omega)=(\omega-{\bf H})^{-1},
\end{equation}
where
\begin{equation}\label{effH}
{\bf H}=$$ \pmatrix{ \epsilon_1-i\Gamma &  -i\Gamma e^{-i\varphi/2}\cos(\varphi/2)\cr
-i\Gamma e^{i\varphi/2}\cos(\varphi/2) & \epsilon_2-i\Gamma \cr
},$$
\end{equation}
is the effective Hamiltonian of the two level system, following the integration of the leads
states. The width $\Gamma$ is given by $2\pi\rho V^2$, where $\rho$ is the DOS of the
leads.

It is possible to write the transmission
amplitude from left to right in terms of the matrix ${\bf G}$
\begin{equation}\label{tgeneral}
t(\omega)= 2\pi \rho \sum_{i,j=1,2} V^*_{L,i} \left[{\bf G}(\omega)\right]_{i\;j} V_{R,j}.
\end{equation}
Thus, the evaluation of
the transmission amplitude requires knowledge of the
diagonal and off-diagonal components of the matrix ${\bf G}$. On
the other hand, the density of states in the QD involves only the
diagonal components of ${\bf G}$ and is given by
\begin{equation}\label{DOS2}
A(\omega)=-2 {\rm Im}\left[ {\rm tr}\left[{\bf
G}(\omega)\right]\right].
\end{equation}
The calculation of ${\bf G}(\omega)$ is
straightforward and yields
\begin{equation}\label{Green2}
{\bf G}=\frac{1}{D(\omega)}$$ \pmatrix{ \omega +
 {\epsilon}+\Delta/2+i\Gamma &  -i\Gamma
e^{-i\varphi/2}\cos(\varphi/2)\cr -i\Gamma
e^{i\varphi/2}\cos(\varphi/2) & \omega+
{\epsilon}-\Delta/2+i\Gamma \cr },$$
\end{equation}
where the denominator is $D(\omega)=(\omega+ {\epsilon})^2-(\Delta/2)^2+2i\Gamma(\omega+  {\epsilon})-\Gamma^2
\sin^2(\varphi/2)$.

\section{Interference and off-diagonal Green's functions}\label{offdiagonal} 

In this appendix we show, by means of a perturbative expansion in $\Gamma$,
that the contribution of the off-diagonal Green's functions
to the transmission amplitude is crucial 
in obtaining a zero in $t_{+}$ and therefore an abrupt phase lapse at $T=0$.

Using Eq.~(\ref{tgeneral}) as well as the expression for the Green's
function matrix ${\bf G}$, Eq.~(\ref{Green2}), one can write the
transmission amplitude $t_{+}$ as
\begin{eqnarray}
t_{+}(\omega)&=&t_{\rm diag}(\omega)+t_{\rm off}(\omega),\\
t_{\rm diag}(\omega)&\equiv&\Gamma\;{\bf G}_{1,1}(\omega) +\Gamma\;{\bf G}_{2,2}(\omega),\\
t_{\rm off}(\omega)&\equiv&\Gamma\;{\bf G}_{1,2}(\omega) +\Gamma\;{\bf G}_{2,1}(\omega),
\end{eqnarray}
where $t_{\rm{diag}(\rm{off})}$ consists of the diagonal/(off-diagonal) 
contributions to $t_{+}$.

Let us now expand the two contributions to $t_{+}$
up to second order in $\Gamma$. Setting $\omega=0$
one obtains
\begin{eqnarray}
t_{\rm diag}(0) &\simeq& -\frac{8\epsilon}{\Delta^2-4\epsilon^2}\Gamma-i\frac{8(\Delta^2+4\epsilon^2)}
{(\Delta^2-4\epsilon^2)^2}\Gamma^2,\\
t_{\rm off}(0) &\simeq& i \frac{8}{\Delta^2-4\epsilon^2}\Gamma^2. 
\end{eqnarray}
The expansion of $t_{\rm diag}$ begins with the linear order in $\Gamma$, describing processes
where an electron hops from the leads to the dot and out. This term is real and 
vanishes for $\epsilon=0$. The imaginary part of $t_{\rm diag}$ is determined by 
the $O(\Gamma^2)$ term in the perturbative expansion, describing events where an electron 
hops twice from the lead to the \it same \rm level before being transfered through the dot.
This term is of the same order in $\Gamma$ as the leading term in the expansion of 
$t_{off}$, the latter describing events where an electron hops twice 
from the lead to \it different \rm levels before being transfered through the dot
(e.g., left lead $\rightarrow$ level 1 $\rightarrow$ left lead $\rightarrow$ level2 $\rightarrow$
right lead). 
It turns out that the term $O(\Gamma^2)$ in the perturbative expansion of $t_{\rm off}$
is purely imaginary and cancels exactly against the $O(\Gamma^2)$ term of $t_{\rm diag}$ at $\epsilon=0$.
More generally, once the contribution of the off-diagonal Green's functions is included, 
the cancellations leading to an exact zero of the transmission amplitude appear order by order
in the perturbative expansion in the dot-lead coupling.

\section{Evaluation of the width of the phase lapse at finite temperatures}\label{AppB}

In this appendix we derive derive the expression for the width of the phase lapse
Eq.~(\ref{lapse1}). The phase observed in an AB interference experiment, $\theta$, 
is defined in Eq.(\ref{phase}). Focusing on the case $s=+1$, the transmission amplitude
Eq.(\ref{transm}) can be expressed as
\begin{eqnarray}\label{transminpoles}
t_{+}(\omega)&=&\frac{2\Gamma\;(\omega_{o}+\epsilon)}{(\omega_{o}-\omega_{e})}\;\frac{1}{\omega-\omega_{o}}-
\frac{2\Gamma\;(\omega_{e}+\epsilon)}{(\omega_{o}-\omega_{e})}\;\frac{1}{\omega-\omega_{e}},
\end{eqnarray}
where the poles $\omega_{e,o}$ are given in Eq.~(\ref{poles}).
The \it temperature averaged transmission
amplitude\rm is therefore given by 
\begin{eqnarray}\label{transmtemperature}
&&t_{+}(T)=-\int d\omega f^{\prime}(\omega) t(\omega)\nonumber \\
&&=\frac{2\Gamma\;(\omega_{o}+\epsilon)}{(\omega_{o}-\omega_{e})}\;\left[\frac{1}{2\pi i\;T}
\Psi^{\prime}\left[\frac{1}{2}+\frac{\omega_{o}}{2\pi i\;T}\right]\right]\nonumber\\
&&-\frac{2\Gamma\;(\omega_{e}+\epsilon)}{(\omega_{o}-\omega_{e})}\;\left[\frac{1}{2\pi i\;T}
\Psi^{\prime}\left[\frac{1}{2}+\frac{\omega_{e}}{2\pi i\;T}\right]\right],
\end{eqnarray}
where $\Psi^{\prime}$ is the trigamma function. In order to obtain the low 
temperature behavior of the transmission phase close to the phase lapse, 
one has to expand Eq.(\ref{transmtemperature}) 
up to second order in $\epsilon$ and $T$ obtaining   
\begin{eqnarray}
t_{+}(T)&=&-\frac{8\Gamma}{\Delta^2} {\epsilon}-i\frac{64 \Gamma^2 (\pi^2 T^2+3\epsilon^2) }{3
\Delta^4} \nonumber \\
&&\simeq -\frac{8\Gamma}{\Delta^2} {\epsilon}-i\frac{64 \Gamma^2 \pi^2}{3
\Delta^4} T^2,
\end{eqnarray}
the last equality being valid for $\mid\epsilon\mid \ll T$.
The width of the phase lapse, $\lambda$, in Eq.~(\ref{lapse1}) follows.
It is obtained straightforwardly from the expression for the temperature
averaged transmission phase
\begin{eqnarray}
\theta(T)&=&{\rm ArcTan}\left(-\frac{{\rm Re}[t]}{{\rm Im}[t]}\right)-\frac{\pi}{2} \nonumber \\
&&= {\rm ArcTan}\left(-\frac{
{\epsilon}}{\lambda}\right)-\frac{\pi}{2}.
\end{eqnarray}

\widetext

\begin{references}
\bibitem{noteYuval} We note that though for $\Gamma >\Delta/2$ 
both $\mid t_{-} \mid^2$ and $A_{+}$ have a single peak at $\epsilon=0$,
in one case the peak height diminishes as $\Gamma$ is further
increased ($\mid t_{-} \mid^2$)
while in the other case the central peak becomes more and more pronounced
($A_{+}$).
\bibitem{noteint} Note that for $s=-1$ and in the presence of local Coulomb
interaction it is possible, by means of change-of-variables
$c_{k,\uparrow}=(c_{k,L}+c_{k,R})/\sqrt{2}$,
$c_{k,\downarrow}=(c_{k,L}-c_{k,R})/\sqrt{2}$, to map our model
onto a pseudo-spin Anderson model\cite{Hofstatter}. Here, the
indices $\uparrow,\downarrow$ are the pseudo-spin states. In this
language, the level spacing $\Delta$ represents  Zeeman splitting.
It is known that this model could give rise to the Kondo effect,
provided $\Delta < T_K$ ($T_K \ll \Gamma$), where $T_K$ is the
Kondo temperature. It follows that in this regime of the model the
Hartree approximation is inappropriate since it 
erroneously breaks the ${\rm SU}(2)$ symmetry. For $\Delta> \Gamma$
(hence $\Delta \gg T_K$) the pseudo-spin ${\rm SU}(2)$ symmetry is
broken, the Kondo effect is suppressed, and employment of the
Hartree approximation is valid.
\end{references}
\end{document}